\documentclass[11pt]{article}
\pdfoutput=1
\usepackage{amsbsy,amsmath,amsthm,amssymb,graphicx,verbatim}
\usepackage{setspace, float, subcaption,xcolor}
\usepackage[longnamesfirst,sort]{natbib}

\usepackage{multirow}

\usepackage{geometry}
\usepackage{algorithm}
\usepackage{algorithmic}

\usepackage{verbatim} 

\theoremstyle{remark}

\newtheorem{example}{Example}

\newfont{\msbm}{msbm10 at 11pt}

\begin{document}
\onehalfspacing

\title{Implementing MCMC: Multivariate estimation with confidence}

\author{James M. Flegal \\ Department of Statistics \\ University of California, Riverside \\ {\tt jflegal@ucr.edu} \and Rebecca P.\ Kurtz-Garcia \\ Department of Mathematical Science and Department of Statistical \& Data Sciences \\ Smith College \\ {\tt rkurtzgarcia@smith.edu}}   

\date{\today}

\maketitle

\begin{abstract}
This paper addresses the key challenge of estimating the asymptotic covariance associated with the Markov chain central limit theorem, which is essential for visualizing and terminating Markov Chain Monte Carlo (MCMC) simulations. We focus on summarizing batching, spectral, and initial sequence covariance estimation techniques. We emphasize practical recommendations for modern MCMC simulations, where positive correlation is common and leads to negatively biased covariance estimates. Our discussion is centered on computationally efficient methods that remain viable even when the number of iterations is large, offering insights into improving the reliability and accuracy of MCMC output in such scenarios.

\end{abstract}

\section{Introduction} \label{sec:intro}

Since the groundbreaking publication by \cite{gelf:smit:1990} Markov Chain Monte Carlo (MCMC) has played a crucial role in the utilization of Bayesian and frequentist statistical models. Usually, an MCMC algorithm generates a sequence of correlated observations to estimate multiple unknown quantities with respect to a target distribution, which could include expectations, density functions, quantiles, modes, and more. Subsequently, the estimated quantities can be employed for inferential purposes.  The main issues we address visualizing and terminating an MCMC simulation; and summarizing batching, spectral, and initial sequence covariance estimation techniques.  Our emphasis is on providing recommendations for modern practical MCMC simulations where positive correlation is routine causing the estimated covariance to be negatively biased.  Moreover, we focus on computationally viable techniques that are appropriate when the number of iterations is substantial. 

Let $F$ be a target probability distribution with support $\mathsf{X} \in \mathbb{R}^{d}$, $d\ge 1$.  A common goal of an MCMC simulation is estimation of several unknown features of $F$.  For example, suppose $g: \mathsf{X}\rightarrow \mathbb{R}^{p}$ be an $F$-integrable function and we are interested in estimating
\[
\theta = E_{F} g(X) = \int_{\mathsf{X}} g(x) F(dx).
\]
The $p$-dimensional vector $\theta$ could include moments, probabilities, or other features of $F$.  For simplicity, we focus on features that can be expressed as expectations but will provide some discussions outside this setting where appropriate (such as quantiles).  

Let $X=\{X_{t}, t\geq 1\}$ be a Harris ergodic Markov chain with invariant distribution $F$.  We defer to other chapters in this volume as to how to produce such a Markov chain \citep[see also][]{robe:case:2004}.  Moreover, we assume a reasonable starting value, or starting distribution, has been identified \citep[see e.g.][]{vats:etal:2020:analyzing, geye:2011}.  Then the ergodic theorem ensures we can estimate $\theta$ from a Monte Carlo sample of size $n$ with the sample mean.  That is, if $Y_{i}=g(X_{i})$ for $i \ge 1$, 
\begin{equation}
\label{eqn:slln}
\bar{\theta}_{n} = \frac{1}{n} \sum_{i=1}^{n} Y_{i}  \stackrel{a.s.}{\to} \theta \text{ as } n\rightarrow\infty .
\end{equation}
Note that if \eqref{eqn:slln} holds for any initial distribution then it holds for every initial distribution \citep{meyn:twee:2009}.  

Naturally there exists an unknown estimation error, $\bar{\theta}_{n} - \theta$, referred to as the Monte Carlo error.  An approximate sampling distribution for the Monte Carlo error can be obtained via a Markov chain central limit theorem (CLT).  An interested reader is directed to \cite{jone:2004} for sufficient conditions for a Markov chain CLT and \cite{jone:hobe:2001} for an introduction on how to establish these conditions.  We assume throughout such a CLT holds.  That is, we assume there exists a $p \times p$ positive definite symmetric matrix $\Sigma$ such that 
\begin{equation}
\label{eqn:clt}
  \sqrt{n} \left( \bar{\theta}_n - \theta \right) \stackrel{d}{\to} \text{N}_{p} \left( 0, \Sigma \right) \text{ as } n\rightarrow\infty,
\end{equation}
where
\begin{equation}
\label{eqn:sigma}
\Sigma = \sum_{k=-\infty}^{\infty}  R(k)
\end{equation}
and $R(k) = \text{Cov}_{F} \left[ Y_{1}, Y_{1+k} \right]$ is the lag-$k$ covariance matrix.  Similar to the ergodic theorem, if \eqref{eqn:clt} holds for any initial distribution then it holds for every initial distribution \citep{meyn:twee:2009}.

The matrix $\Sigma$ is referred to as the asymptotic covariance associate with a Markov chain CLT, or more simply as the long run variance (LRV).  Since $\Sigma$ is positive definite there is exactly one positive definite matrix $B$ such that  $\Sigma = B^{T} B$, where $B$ is the positive square root.  Moreover, Monte Carlo standard errors (MCSEs) for each component of $\theta$ can be obtained by estimating the vector of length $p$ given by $\text{diag} (B)/\sqrt{n}$.  Hence, one can use the CLT at \eqref{eqn:clt} to access variability of the estimator $\bar{\theta}_{n}$ provided an estimate of $\Sigma$, say $\Sigma_n$, is available.  We explore three broad classes of estimators appropriate for this task.  

The rest of the chapter is organized as follows.  Section~\ref{sec:output} describes a number of multivariate MCMC output analysis tools based on \eqref{eqn:clt} that require high-quality estimates of $\Sigma$.  Section~\ref{sec:batch} considers batching estimators for $\Sigma$ and introduces the lugsail transformation. Section~\ref{sec:spectral} summarizes spectral variance (SV) estimators for $\Sigma$ and the role lag windows play in their construction.  Section~\ref{sec:initial} introduces initial sequence estimators for $\Sigma$, which leverage the properties of the sequence of lag-$k$ covariance matrices leading to conservative estimates. Section~\ref{sec:examples} illustrates these variance estimation techniques using a Bayesian logistic regression for modeling credit risk that specifically compares finite sample performance and computational time.  We conclude with a discussion in Section~\ref{sec:discussion}.

\section{MCMC output analysis} \label{sec:output}

A crucial step in a MCMC experiment is considering the Monte Carlo error resulting from finite simulation.  In short, it's advisable to include individual MCSEs or marginal confidence intervals to assess the variability of the estimator(s) $\bar{\theta}_n$, as emphasized by \cite{fleg:hara:jone:2008}.  Additionally, practitioners can use the Monte Carlo error to help determine the simulation's duration.

Practitioners can address these aspects in conjunction by employing a sequential stopping rule that terminates the simulation when the confidence interval width for the estimator(s) of interest reaches a user-defined threshold, as discussed by \cite{jone:hara:caff:neat:2006}.  Alternatively, an extension of this rule can be used which terminates the simulation when computational uncertainty becomes sufficiently small compared to posterior uncertainty. Specifically, \cite{fleg:gong:2015} propose a relative sequential stopping rule that terminates when the confidence interval width is sufficiently small relative to the posterior standard deviation of the target parameter.  \cite{gong:fleg:2016} establish that this approach is equivalent to stopping when the effective sample size (ESS) is adequately large.

Both methods critically rely on estimates for the diagonal entries of $\Sigma$, which are used to estimate of $\text{diag} (B)/\sqrt{n}$.  Recent advancements in MCMC output analysis incorporate correlation information from the off-diagonal elements of $\Sigma$. We next provide an overview of such an approach and relevant features, underlining the necessity for accurate multivariate estimation techniques of $\Sigma$.

\subsection{Multivariate sequential stopping rule}

For simplicity, consider the $p$-dimensional vector of sample means $\bar{\theta}_n$ in \eqref{eqn:slln}.  Suppose $\chi^{2}_{1-\alpha, p}$ denotes a $1-\alpha$ quantile from a $\chi^2$ distribution with $p$ degrees of freedom and $| \cdot |$ is the determinant.  Then an asymptotic $100(1-\alpha)\%$ confidence region can be constructed as
\[ 
C_{\alpha}(n) = \left\{ n(\bar{\theta}_n - \theta)^T \Sigma_{n}^{-1}  (\bar{\theta}_n - \theta) < \chi^{2}_{1-\alpha, p} \right\}  . 
\]
The $p$-dimensional ellipsoid $C_{\alpha}(n)$ has volume
\[
\text{Vol} \left( C_{\alpha}(n) \right) = \frac{ 2 \pi^{p/2} }{ p \Gamma(p/2) } \left( \frac{ \chi^{2}_{1-\alpha, p} }{ n } \right)^{p/2} \left| \Sigma_n \right| ^{1/2} \; ,
\]
which can be used to measure the variability of the estimator $\bar{\theta}_{n}$.   

Practitioners can terminate the simulation when the computational uncertainty is small relative to the model uncertainty in multivariate settings.  To this end, suppose the model uncertainty is measured using the generalized variance of the target $F$, i.e.\ $|\Lambda|$ where $\Lambda = \text{Var}_F(Y_1)$.  We can estimate $\Lambda$ with the sample covariance matrix, denoted as $\Lambda_n$.  A relative standard deviation sequential fixed-volume stopping rule,  proposed by \cite{vats:fleg:jone:2019}, terminates the simulation when the ellipsoid volume is sufficiently small relative to the volume of $\Lambda_n$.  That is, the simulation terminates when the volume of $C_{\alpha}(n)$ is an $\epsilon^{ith}$ fraction of the size of $\Lambda_n$ for some desired tolerance level $\epsilon > 0$.  Formally, the fixed-volume stopping rule terminates the first time
\begin{equation} \label{eqn:fixedvolume}
\left\{ \text{Vol} \left( C_{\alpha}(n) \right) ^{1/p} + 1/n < \epsilon |\Lambda_n| ^{1/2p} \right\} ,
\end{equation}
for $n > n^* > 0$.  The role of $n^*$ is to ensure stable estimates for $\Lambda$ and $\Sigma$ have been obtained.  While there is no theoretical restrictions on $n^*$, we elaborate on a reasonable choice later.

The fixed-volume stopping rule is asymptotically equivalent to simulating until a (multivariate) ESS is sufficiently large.  Specifically, ESS measures the sample size required to achieve the same generalized Monte Carlo error if the sample was from an independent and identically distributed sequence, defined as
\[
\text{ESS} = n \left(\dfrac{|\Lambda|}{|\Sigma|} \right)^{1/p}\, .
\]
Since ESS is unknown, it is estimated using 
\begin{equation} \label{eqn:ESS}
\widehat{\text{ESS}} = n \left(\dfrac{|\Lambda_n|}{|\Sigma_n|} \right)^{1/p}\,.
\end{equation}
When $p=1$, \eqref{eqn:ESS} reduces to the typical estimate of univariate ESS \citep{robe:case:2004, kass:1998}.  Since the ratio of generalized variances in \eqref{eqn:ESS} also appears in \eqref{eqn:fixedvolume}, the fixed-volume stopping rule can be shown to be asymptotically equivalent terminating when 
\[
\widehat{\text{ESS}} \ge M_{\alpha, \epsilon, p} ,
\]
where $M_{\alpha, \epsilon, p}$ depends only on the confidence level, relative precision desired, and dimension of the estimation problem.  Hence $M_{\alpha, \epsilon, p}$ can be calculated prior to simulation, which is available via the \texttt{minESS} function \citep{mcmcse2021}.  For example, suppose we are interested terminating the simulation when the volume of a 95\% confidence region ($\alpha = 0.05$) is 0.05 the volume of a posterior covariance matrix $\Lambda$.  Then dimensions $p$ of 1, 3, or 10 would yield $M_{\alpha, \epsilon, p}$ of 6146, 8123, or 8831, respectively.  The value $M_{\alpha, \epsilon, p}$ provides a reasonable choice for a minimum simulation size $n^*$.  

\cite{vats:knudson:2021} establish a relationship between ESS and the Gelman-Rubin-Brooks diagnostic \citep{broo:gelm:1998, gelm:rubi:1992a}, which is also used as a stopping criteria.  However, we recommend using fixed-volume or ESS stopping rules since they are easily interpretable, have lower variability, and are readily available in software. 

\begin{figure}
\begin{center}
 {\includegraphics[height=3in]{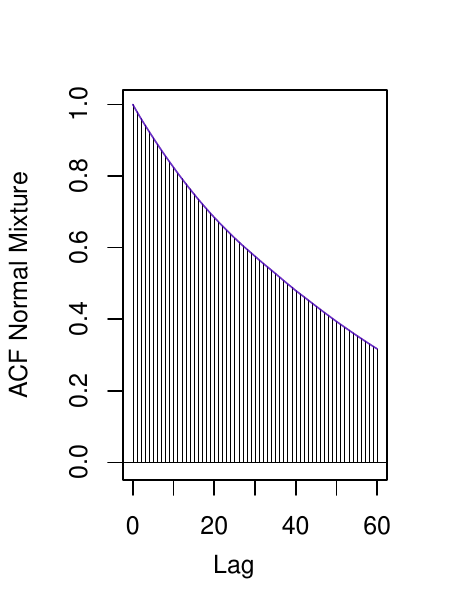}} 
 {\includegraphics[height=3in]{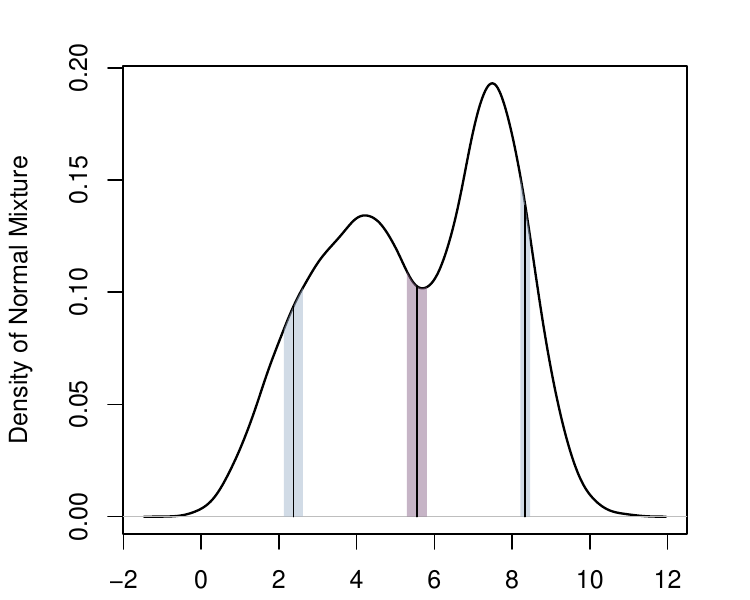}} 
 \caption{\label{fig:SimToolsPlot}Normal mixture density ACF plot and nonparametric density plot with simultaneous error bars surrounding estimates of the mean and endpoints of an 80\% credible interval.}
\end{center}
\end{figure}

\begin{example}
Consider the three-component normal mixture density
\begin{equation}\label{ex_Mixed_Normal_Distribution_1}
f(x) = .2 \phi(x; 2.5, 1^2) + .3 \phi(x; 4.5, 1^2) + .5 \phi(x; 7.5, 1^2), 
\end{equation}
where $\phi(x; m, s^2)$ denotes a normal density function with mean $m$ and variance $s^2$.  We simulate $5e4$ correlated samples from this density via a random walk Metropolis-Hasting algorithm with proposal standard deviation of $1/2$.  

The left panel of Figure~\ref{fig:SimToolsPlot} plots the correlation coefficient against the lag where the observed lag 1 autocorrelation is approximately 0.98.  While this high level of correlation suggest a larger proposal variance should be considered, we continue using this Metropolis-Hastings update for illustrative purposes.  Consider estimating $E_F X$, which we know to be 5.6 from \eqref{ex_Mixed_Normal_Distribution_1}, using $\bar{x} = 5.5493$ with an MCSE of 0.11549.  The ESS based on estimation of the mean can also be estimated as $\widehat{\text{ESS}} = 386$.  Since this is lower than $M_{.05, .10, 1} = 1536$, it appears the simulation should continue to ensure higher accuracy in estimation of the mean.  
\end{example}

\subsection{Visualizing simultaneous simulation error}

Up to this point we have only considered expectations, however estimation of quantiles associated with $F$ are often of interest.  For example when $F$ is a posterior, the quantile of interest could be from a marginal posterior distribution.  Let $h:\mathcal{X} \to \mathbb{R}$ and $V = h(X)$ with distribution function $F_{h}(v)$.  Further assume $F_{h}(v)$ is absolutely continuous with a continuous density $f_{h}(v)$.  Then define the $q$-quantile associated with $F_{h}$ as
\begin{equation*}
\xi_{q} = F_{h}^{-1}(q) = \inf\{ v:F_{h}(v)\geq q\}.
\end{equation*}
The quantile $\xi_{q}$ can be estimated using $\hat{\xi}_{q} = h (X)_{\lceil nq \rceil:n}$, which is the $\lceil nq\rceil^{th}$ order statistic of $h(X)$.  \cite{doss:fleg:jone:neat:2014} establish conditions under which the sampling distribution of the Monte Carlo error is approximately normal and provide techniques to estimate the associated LRV.  

Estimation on a finite combination of $p$ expectations and quantiles can also lead to a joint sampling distribution.  An interested reader is directed to \cite{robertson2021assessing} for conditions that ensure such a CLT holds and practical techniques to estimate the associated $p \times p$ positive definite LRV $\Omega$.  In short, define the $p$-vector $\nu$ containing the expectations and quantiles that we aim to estimate with the $p$-vector $\hat{\nu}$ the corresponding ergodic averages and order statistics, then as as $n \to \infty$, we assume
\begin{equation} 
\label{eq:CLT.joint}
\sqrt{n} \left( \hat{\nu} - \nu \right) \overset{d}\to N_{p} \left(0, \Omega \right) .
\end{equation}

After simulating $n$ MCMC iterations, practitioners may want to visualize the results of MCMC experiments and corresponding variability.  However, visualizing a $p$-dimensional ellipsoid is difficult beyond two dimensions and plots of multiple marginal intervals can be difficult to interpret.  To this end, \cite{robertson2021assessing} also provide $p$-dimensional simultaneous confidence regions for $\nu$ using \eqref{eq:CLT.joint} and a strongly consistent estimator of $\Omega$, say $\hat{\Omega}$.  Their approach considers hyperrectangular regions of the form
\begin{equation} 
\label{eq:C.SI}
  C_{SI}(z) = \prod_{i=1}^{p} \left[ \hat{\nu}_i - z \dfrac{\hat{\Omega}_{i,i}}{n}, \hat{\nu}_i + z \dfrac{\hat{\Omega}_{i,i}}{n} \right]\,,
\end{equation}
where $z > 0$.  

Then they propose a parametric approach to find $z^*$ such that $C_{SI}(z^*)$ has a user-defined simultaneous coverage probability $1-\alpha$.  Specifically, suppose $U \sim N_p(\hat{\nu}, \hat{\Omega}/n)$ and notice that $\Pr(U \in \mathcal{C}_{SI}(z))$ is strictly increasing as $z > 0$ increases.  Then a univariate optimization can be implemented to find $z^*$ such that $\Pr(U \in \mathcal{C}_{SI}(z^*)) \approx (1-\alpha)$ where the multivariate normal probabilities can be quickly and accurately calculated using quasi-Monte Carlo methods \citep{R:mvtnorm}.  Finally, the marginal intervals at \eqref{eq:C.SI} can be incorporated into standard plots enabling practitioners to more easily assess the reliability of an MCMC simulation through visual tools.  The following example illustrates one such visualization of the simultaneous simulation error.

\begin{example}
Recall the three-component normal mixture density at \eqref{ex_Mixed_Normal_Distribution_1} where $5e4$ correlated samples were obtained from a random walk Metropolis-Hasting algorithm.  We now consider simultaneous estimation of the mean $E_F X$ and an 80\% credible interval for $X$ with endpoints $\xi_{.10}$ and $\xi_{.90}$, where under regularity conditions as $n \to \infty$,
\begin{equation*} 
\label{eq:CLT.ex}
\sqrt{n} \left( 
\begin{pmatrix} \bar{x}_n \\ \hat{\xi}_{.10} \\ \hat{\xi}_{.90} \end{pmatrix} - 
\begin{pmatrix} E_F X \\ \xi_{.10} \\ \xi_{.90} \end{pmatrix}
\right) \overset{d}\to N_{3} \left(0, \Omega \right) .
\end{equation*}
The right panel of Figure~\ref{fig:SimToolsPlot} plots the estimated mean and 80\%  credible interval along with a nonparametric density estimate of $f(x)$.  The error bars around the mean and quantiles account for the correlated nature of the process and have a simultaneous nominal 0.95 coverage probability.
\end{example}

The accuracy of inference and estimated ESS in the preceding example are significantly affected by the LRV estimation quality.  If the LRV is underestimated, it can lead to smaller than appropriate MCSEs, which in turn results in an inflated ESS and excessively narrow simultaneous (or univariate) error bars. The following sections introduce three distinct techniques for estimating $\Sigma$. These techniques are particularly valuable in scenarios with large sample sizes and high correlations, which are common in MCMC simulations.

\section{Batching methods} \label{sec:batch}

In MCMC simulations the batch means (BM) estimator for $\Sigma$ is most commonly implemented since it's easy to use and computationally efficient \citep{chen:seila:1987, geye:1992}.  The BM estimator uses $a$ sequential non-overlapping batches of length $b$ to estimate $\Sigma$.  Suppose $n = a b$ and let $\bar{Y}_k= b^{-1}\sum_{i=1}^{b} Y_{kb + i}$ for $k = 0, \dots, a-1$.  Then the BM
estimator with batch size $b$ is,
\begin{equation} \label{eq:bm}
\hat{\Sigma}_{n,b}= \frac{b}{a-1} \sum_{k = 0}^{a-1} \left(\bar{Y}_k - \bar{\theta}_n\right) \left(\bar{Y}_k - \bar{\theta}_n \right)^T\,.
\end{equation}
The BM estimator has been well studied in MCMC settings \citep{fleg:jone:2010, jone:hara:caff:neat:2006, vats:fleg:jone:2019, chakraborty2022estimating}.  However, BM estimators suffer from negative bias in finite samples when the Markov chain has positive correlation.  A careful discussion of the bias requires defining 
\[
\Gamma = - \sum_{k=-\infty}^{\infty} k R(k) \,.
\]
Then the bias of the BM estimator can be obtained, i.e.
\[
\text{\text{Bias}} \left(\hat{\Sigma}_{n, b} \right) =  \dfrac{\Gamma}{b} + o \left(\dfrac{1}{b} \right)\,.
\]
The first-order bias term, $\Gamma / b$, has negative diagonals for positively correlated processes.  Moreover the magnitude of the negative bias grows as the correlation approaches 1.  

To address this bias, \cite{vats2022lugsail} propose a linear combination of two BM estimators with batch sizes $b$ and $\lfloor b/r \rfloor$ for $r \ge 1$.  Specifically, if $c_n \in [0, 1)$ then the lugsail BM estimator is
\begin{equation}
\label{eq:lugsail_bm}
	\hat{\Sigma}_{n, L} = \dfrac{1}{1-c_n} \hat{\Sigma}_{n, b} - \dfrac{c_n}{1-c_n}\hat{\Sigma}_{n, \lfloor b/r \rfloor }\,.
\end{equation}
When $r = 1/c_n$ the estimator \eqref{eq:lugsail_bm} has a first-order bias of 0, which has been referred to as the zero lugsail BM estimator or flat-top BM estimator \citep{liu:fleg:2018}.  More generally, \cite{vats2022lugsail} show
\begin{equation}
\label{eq:bias_lugBM}
\text{\text{Bias}}\left(\hat{\Sigma}_{n, L}\right) =  \dfrac{\Gamma}{b} \left(\dfrac{1 -rc_n}{1-c_n}\right) + o\left( \dfrac{1}{b}\right)\,.
\end{equation}
Hence, setting $r > 1/c_n$ can induce a positive first-order bias for positively correlated chains.

\cite{vats2022lugsail} provide recommendations for $r$ and $c_n$ based on estimating the underlying lag 1 autocorrelation of the Markov chain.  When the underlying correlation is low, $\rho \in [0, .7)$, the authors suggest setting $r=2$ and $c_n = 1/2$ resulting in a zero lugsail BM estimator.  For moderate correlation settings, $\rho \in [.7, .95)$, they suggest an adaptive lugsail where $r=2$  and $c_n$ varies based on $n$ but ultimately converges to the zero lugsail BM estimator.  Specifically, 
\[
c_n = \dfrac{\log(n) - \log(b) + 1}{2(\log(n) -\log(b)) + 1}\,.
\]
For high correlation settings, $\rho \in [.95, 1)$, they suggest setting $r=3$ and $c_n = 1/2$ resulting in an over lugsail BM estimator.  The first-order bias for the over lugsail setting is intentionally over corrected to account for higher-order and finite sampling biases that also lead to underestimating $\Sigma$ in practice.  Since the over lugsail tends to overestimate $\Sigma$ while remaining asymptotically unbiased, it is particularly useful in conjunction with sequential stopping rules.   A downside of lugsail estimators is that they have a higher variance than the standard BM estimator for fixed $b$.

The following example illustrates the utility of lugsail estimators with a univariate data set.  Calculations in the example utilize the exact bias expression of \cite{akta:tuba:2007} for the univariate BM estimator, i.e.\ when $p=1$ 
\[
\text{Bias} \left( \hat{\Sigma}_{n,b} \right)
= -\dfrac{2(a+1)}{ab} \sum_{s=1}^{b - 1} sR(s) - 2 \sum_{s=b}^{\infty} R(s) -   \dfrac{2}{a-1} \sum_{s = b}^{n-1}\left( 1 - \dfrac{s}{n} \right) R(s) \; .
\]
This result enables exact bias calculations for lugsail estimators due to the linear relationship at \eqref{eq:lugsail_bm}.

\begin{example}
The normal AR(1) times series is given by
\[
X_{n+1} = \phi X_n + \epsilon_n \; ,
\]
where the $\epsilon_n$ are i.i.d.\ $N(0,1)$ and $| \phi | < 1$.  This Markov chain has invariant distribution $N \left( 0, 1 / (1 - \phi^2) \right)$.  When estimating the mean of this distribution, $E_{F} X = 0$, $\Sigma = 1/(1- \phi)^2$ and $\text{ESS} / n = (1- \phi)^2 / (1 - \phi^2)$.  We investigate two scenarios with correlation coefficients, denoted as moderate ($\phi = 0.92$) and high ($\phi = 0.98$), respectively. Our aim is to demonstrate the impact of bias on coverage probability and ESS calculations in finite sample contexts.  For simplicity $b = \lfloor \sqrt{n} \rfloor$ in this example. 

\begin{figure}
\begin{center}
 {\includegraphics[height=3.5in]{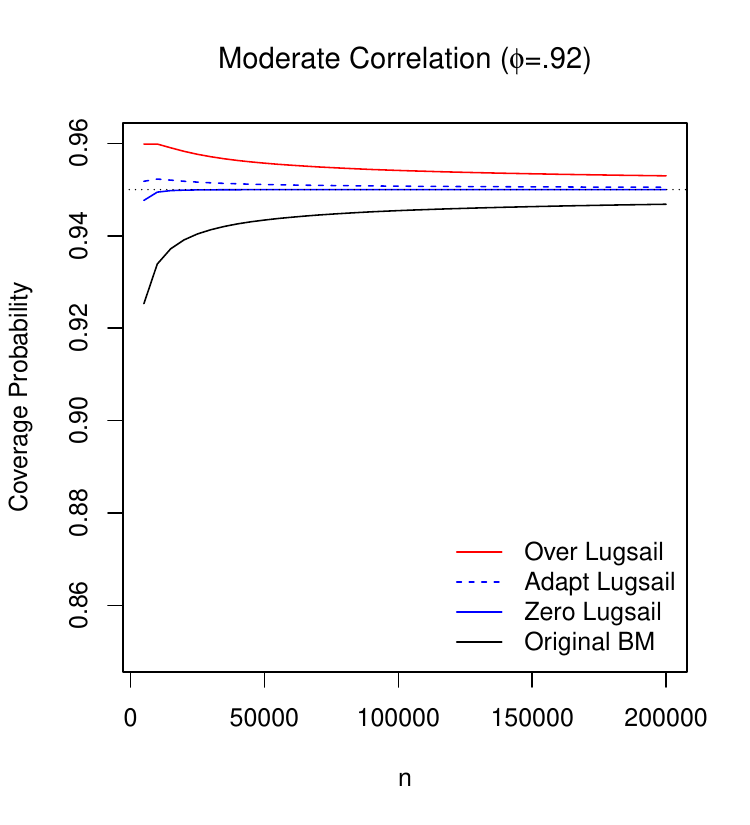}} 
 {\includegraphics[height=3.5in]{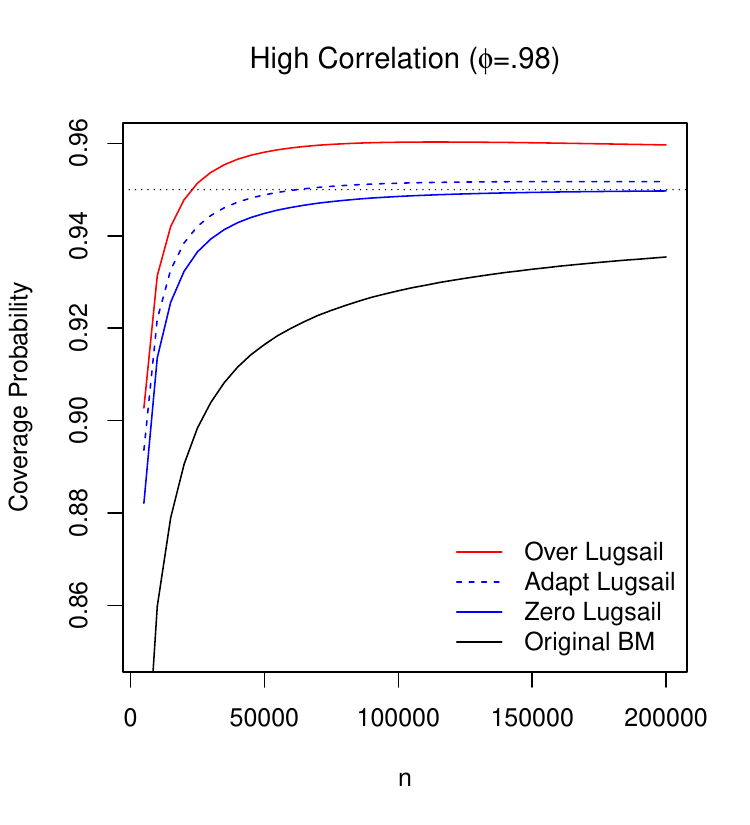}} 
 \caption{\label{fig:true.cov}Coverage probability of a 95\% confidence interval for $E_{F} X$, using various variance estimation methods.}
\end{center}
\end{figure}  

In Figure~\ref{fig:true.cov}, the observed coverage probability of a 95\% confidence interval for $E_{F} X$, using various variance estimation methods, is depicted. The original BM estimator gradually approaches the nominal level of 0.95, suggesting an underestimation of finite sample confidence intervals. This underestimation is notably in both scenarios, where the coverage probability hovers around 0.935 after $2e5$ iterations when $\phi = 0.92$. Although this deviation might seem minor, it becomes magnified in multivariate applications like the simultaneous confidence regions demonstrated in Figure~\ref{fig:SimToolsPlot}.

Both the zero and adapt lugsail estimators similarly tend to converge towards the expected value from below, albeit at a faster rate. The advantage of the over lugsail estimator lies in its behavior of converging from above in both correlation settings, following an initial simulation effort.  This has a statistical advantage in that the observed coverage probability will at least obtain the nominal confidence level of 95\%, satisfying the requirements for a valid Neyman-Pearson inference procedure. 

Consistent findings are drawn from Figure~\ref{fig:true.ess}, where the relationship between simulation size and expected $\text{ESS}/n$ is depicted for different variance estimation techniques. The dotted horizontal black line corresponds to the actual $\text{ESS} / n$, which varies with $\phi$. The advantage of converging from below exhibited by the over lugsail estimator is that it guards against premature termination of simulations. Subsequent examples, where the true value cannot be calculated, will exhibit similar patterns.

\begin{figure}
\begin{center}
 {\includegraphics[height=3.5in]{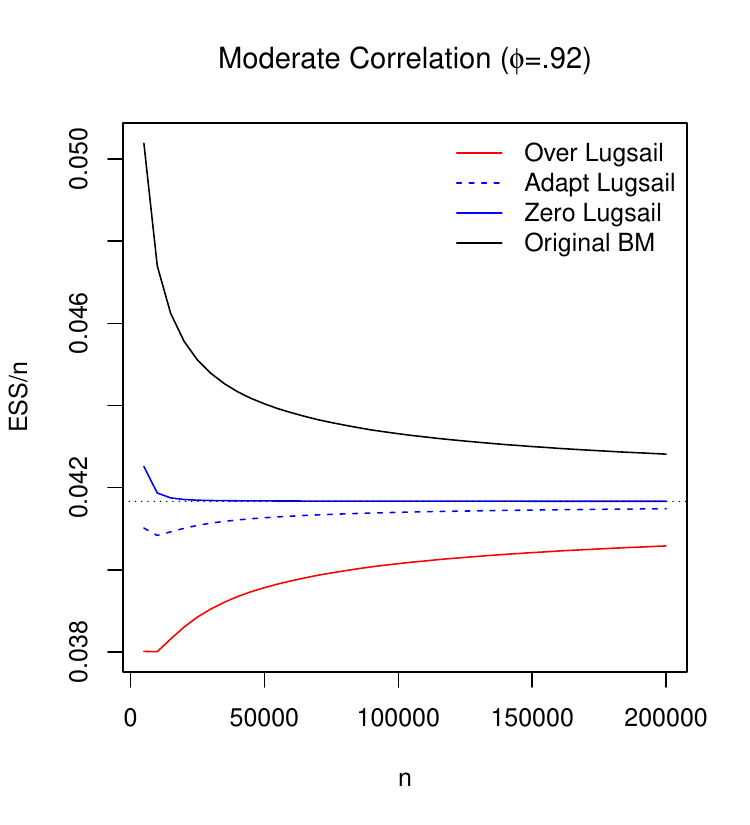}} 
 {\includegraphics[height=3.5in]{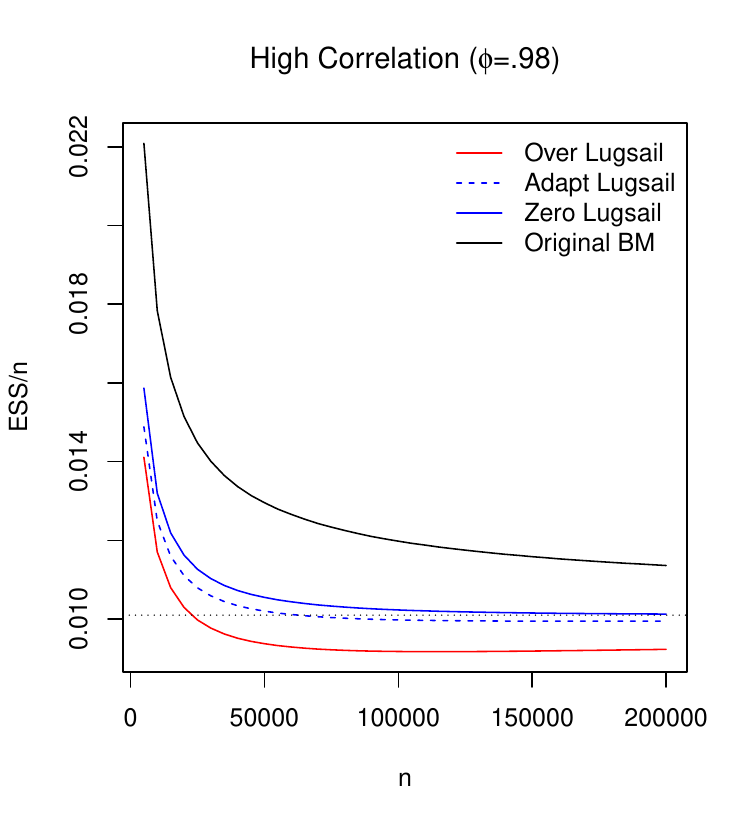}} 
 \caption{\label{fig:true.ess}Plots of simulation size versus $\text{ESS}/n$, using various variance estimation methods.}
\end{center}
\end{figure}  

\end{example}

Choosing the batch size $b$ is critical to the finite sample behaviour of the BM estimator.  The lugsail BM estimator requires that both the number of batches $a \to \infty$ and batch size $b \to \infty$ as $n \to \infty$ to obtain strong and mean-square consistency.  For the BM estimator, \cite{fleg:jone:2010} establish the mean-squared-optimal batch size $b \propto n^{1/3}$, where as \cite{liu2021batch} provides a parametric approach to estimate the proportionality constant.  This proportionality constant is a function of $\Sigma$ and $\Gamma$ and hence depends on the amount of serial correlation in the Markov
chain.  

Batch size selection for the more general lugsail BM estimator remains an open problem.  For example, an optimal batch size for the zero lugsail estimator should be smaller than that of the standard BM estimator.  However, a mean-squared-optimal approach based solely on the first order bias would net an optimal $b$ value that minimizes only the variance by setting $b=0$.  Hence a more nuanced approach is necessary.  Fortunately, current best practices are readily available to practitioners in the \texttt{mcmcse} R package \citep{mcmcse2021}.  

\subsection{Other batch based estimators}

There are a number of additional batch based estimators of $\Sigma$.  Overlapping BM use $n-b+1$ overlapping batches of length $b$ denoted $\dot{Y}_{l}(b)=b^{-1}\sum_{t=1}^{b}Y_{l+t}$ for $l = 0, \dots, n-b$.  Then the overlapping BM estimator is given by
\begin{equation} \label{eq:olbm}
\hat{\Sigma}_{obm}=\dfrac{nb}{(n-b)(n-b+1)}\sum_{l=0}^{n-b}(\dot{Y}_{l}(b)-\bar{Y}_{n})(\dot{Y}_{l}(b)-\bar{Y}_{n})^T.
\end{equation}
The first-order bias of overlapping BM is equivalent to that of the BM estimator but computing overlapping BM is slower given the increased quantity of batches.  One advantage of overlapping BM is a reduction in its variability.  Specifically, the variance of the BM estimator is 1.5 times higher than that of the overlapping BM estimator \citep{fleg:jone:2010}.  

One could consider a linear combination of overlapping BM estimators, similar to \eqref{eq:lugsail_bm}.  This remains an open problem but the first-order bias will remain equal to \eqref{eq:bias_lugBM}.  Other BM variants include a recursive estimator which employs a sequence of batch sizes that increase as $n$ increases \citep{chan:yau:2017} and a replicated estimator for estimating $\Sigma$ from parallel Markov chains \citep{gupta2020estimating, argon2006replicated}.

Lag windows, introduced the following section, can also be incorporated in to batched estimators.  To this end, \cite{dame:1987, dame:1991} and \cite{vats:fleg:jone:2018} study a generalized overlapping BM estimator and establish that it is asymptotically equivalent to SV estimators.   \cite{liu:fleg:2018} consider a multivariate non-overlapping version, which they refer to as weighted BM.  Use of these with the Bartlett lag window leads to the BM and overlapping BM estimators at \eqref{eq:bm} and \eqref{eq:olbm}.  Using the Bartlett flat-top lag window in the weighted BM estimator leads the the zero lugsail BM estimator.  

\section{Spectral methods} \label{sec:spectral}

SV methods can also be used to estimate $\Sigma$ in MCMC simulations \citep{fleg:jone:2010, vats:fleg:jone:2018}.  First consider estimating the lag-$k$ covariance matrix $R(k)$ with the sample lag-$k$ covariance
\begin{equation}
\label{eq:sample_lag_covariance}
\hat{R}(k) = \dfrac{1}{n} \sum_{i=1}^{n-k}\left(Y_i - \bar{\theta}_n \right)\left(Y_{i+s} - \bar{\theta}_n \right)^T\,.
\end{equation}
SV estimators weight the sample lag covariances in \eqref{eq:sample_lag_covariance} using a lag window function $\kappa : \mathbb{R} \to \mathbb{R}$ such that $\kappa(0) = 1, \kappa(x) = \kappa(-x)$ for all $x \in \mathbb{R}$.  Suppose $b \in \mathbb{N}$ is the truncation point, then the multivariate SV estimator is
\begin{equation}
\label{eq:sve}
\dot{\Sigma}_{k,b} = \sum_{s= -(n - 1)}^{n-1} \kappa\left( \dfrac{s}{b}\right) \hat{R}(s)\,.
\end{equation}

\begin{table}
  \caption{\label{tab:lag_windows}Common lag windows.}
\begin{center}
\begin{tabular}{ccc}
\multicolumn{1}{c}{Lag Window} & \multicolumn{1}{c}{$\kappa(x) = $} \\ \hline
\noalign{\vskip 2mm}   
Bartlett     & $\left(1 - |x|\right) I ( |x| \leq 1 )$ \\
\noalign{\vskip 2mm}   
Bartlett Flat-Top & $I \left( |x| \leq \dfrac{1}{2} \right) + 2 \left( 1 - |x| \right) I \left( \dfrac{1}{2} < |x| \leq 1 \right)$ \\
\noalign{\vskip 2mm}   
Tukey-Hanning & $\dfrac{1}{2} + \dfrac{1}{2} \cos(\pi x) I ( |x| \leq 1 )$ \\
\noalign{\vskip 2mm}   
Quadratic Spectral & $\dfrac{25 }{12 \pi^2 x^2} \left\{\dfrac{\sin \left(6 \pi x /5 \right) }{6 \pi x /5}  - \cos\left( 6 \pi x /5 \right)\right\} $ \\
\noalign{\vskip 2mm}   
\hline
\end{tabular}
\end{center}
\end{table}

\begin{figure}
\begin{center}
 {\includegraphics[height=3.5in]{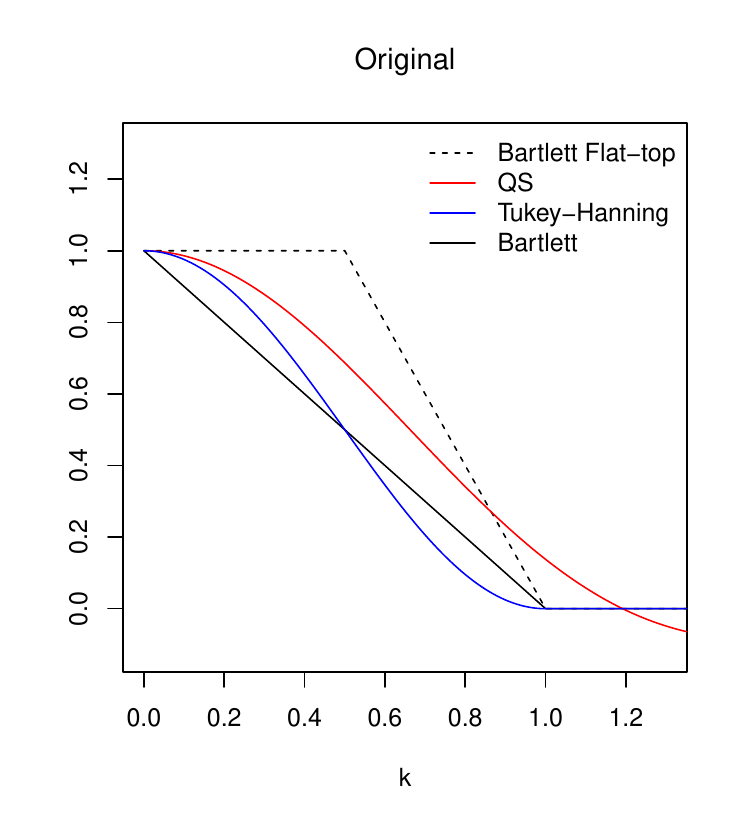}} 
 \caption{\label{fig:original}Plots of original lag windows.}
\end{center}
\end{figure}

Table~\ref{tab:lag_windows} provides some common lag windows, which are also plotted in Figure~\ref{fig:original}.  The Bartlett lag-window is the most popular and routinely referred to as the \cite{newey:west:1987} estimator in econometrics.  It is well known the overlapping BM estimator at \eqref{eq:olbm} is asymptotically equal to the SV estimator with a Bartlett lag window apart from some end effects \citep[see e.g.][]{welc:1987, meke:schm:1984, vats:fleg:jone:2019}.

Figure~\ref{fig:original} illustrates Bartlett, Tukey-Hanning, and Quadratic Spectral lag windows decrease for $x \in \left[ 0, \epsilon \right)$ for some $\epsilon >0$, which leads to underestimation of $\Sigma$ for positively correlated sequences.  The Bartlett flat-top lag window \citep{poli:roma:1995,poli:roma:1996} has a slope of zero at $x=0$ and as a result has a first-order bias of zero.  

The general first-order bias expression depends on the smoothness of the kernel at zero, and requires more notation.  To this end, define
\[
k_q = \lim_{x \to 0} \dfrac{1 - k(x)}{|x|^q}
\]  
for $q \in [0, \infty)$. If we let $q \ge 1$ be the largest integer for which $k_q < \infty$ we will observe that smoother kernels near 0 result larger values of $q$, see \cite{andr:1991} and \cite{parzen:1957} for additional discussion. A key object in a general first-order bias expression is 
\[
\Gamma^{(q)} = - \sum_{k=-\infty}^{\infty} k^q R(k) \; ,
\]
which generalizes $\Gamma$.  Then the bias of the SV estimator can be obtained, i.e.
\[
\text{\text{Bias}} \left( \dot{\Sigma}_{k,b} \right) =  \dfrac{k_q \Gamma^{(q)}}{b^q} + o \left(\dfrac{1}{b^q} \right)\,.
\]
Again, the first-order bias term, $k_q \Gamma^{(q)} / b^q$, has negative diagonals for positively correlated processes and the magnitude of the negative bias grows as the correlation approaches 1.  

\begin{figure}
\begin{center}
 {\includegraphics[height=3.5in]{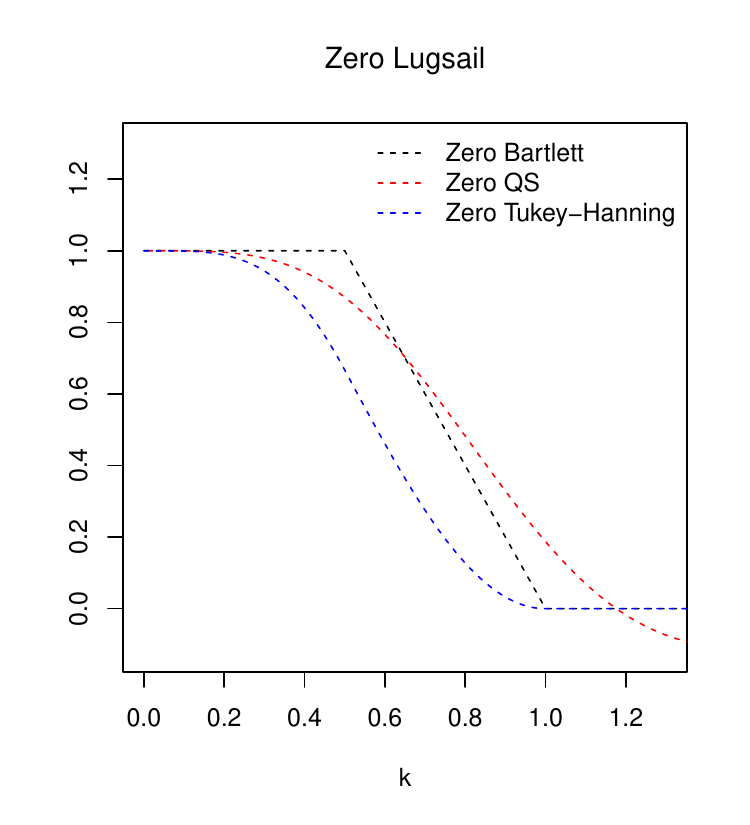}} 
 {\includegraphics[height=3.5in]{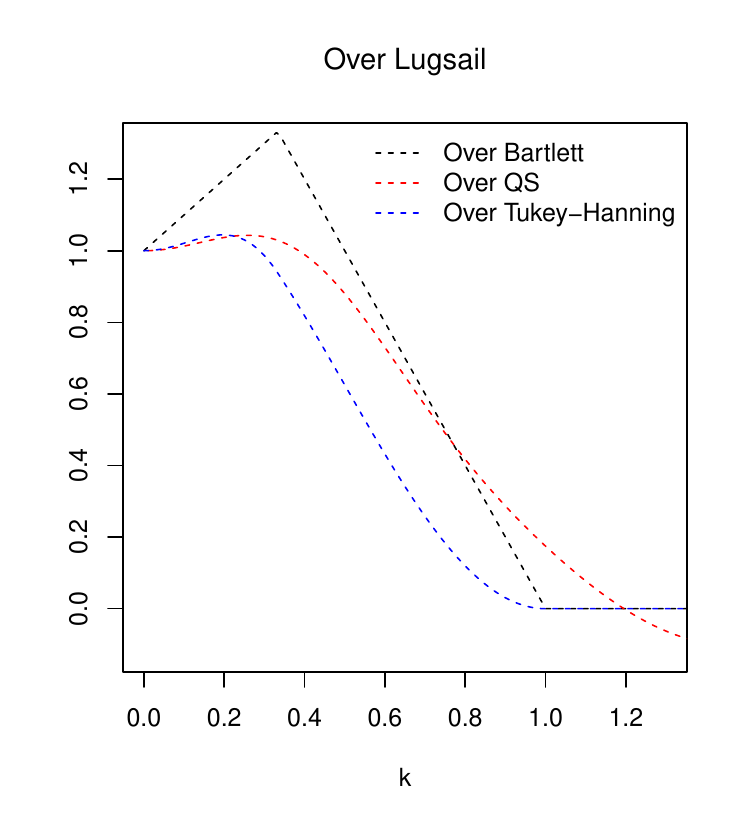}} 
 \caption{\label{fig:lugsail}Plots of zero and over lugsail versions of the Bartlett, Tukey-Hanning, and Quadratic Spectral lag windows.}
\end{center}
\end{figure}

Lugsail lag windows of \cite{vats2022lugsail} generalize traditional lag windows and allow settings that lift the weights above 1 to induce a positive first-order bias.  Suppose $r \ge 1$ and $ c_n\in [0,1)$ is a sequence such that $c_n \to c$ as $n \to \infty$, then the family of lugsail windows associated any existing lag window $\kappa$ is defined as
\begin{equation}
\label{eq:lugsail}
\kappa_L(x) = \dfrac{1}{1-c_n} \kappa(x) - \dfrac{c_n}{1-c_n} \kappa(rx)\,.
\end{equation}
Setting $c_n = 0$ or $r = 1$ results in the original lag window.  Using the Bartlett lag window with $r=2$ and $c_n = 2$ gives the Bartlett flat-top lag window in Table~\ref{tab:lag_windows}.  The left panel of Figure~\ref{fig:lugsail} shows zero lugsail versions of the Bartlett, Tukey-Hanning, and Quadratic Spectral lag windows, which have a first-order bias of zero.  The right panel of Figure~\ref{fig:lugsail} shows over lugsail versions of the same three lag windows, which have a positive first-order bias in a effort to offset most of the (unknown) high-order bias terms.  

The lugsail lag window with the multivariate SV estimator can alternatively be characterized as a linear combination of SV estimators with truncation points $b$ and $\lfloor b/r \rfloor$.  That is, the lugsail SV estimator is
\begin{equation} \label{eq:sv.linear}
\dot{\Sigma}_{k,L} =  \dfrac{1}{1-c_n}\dot{\Sigma}_{k,b} - \dfrac{c_n}{1-c_n}\dot{\Sigma}_{k, b/r }\, .
\end{equation}
From this linear relationship we observe that the lugsail estimators retain weak and strong consistency from the original SV estimators, sufficient conditions can be found in \cite{vats:fleg:jone:2018}.  \cite{vats2022lugsail} establish the first-order bias expression for lugsail lag windows
\begin{equation*}
\label{eq:bias_lugSV}
\text{\text{Bias}}\left( \dot{\Sigma}_{k,L} \right) =  \dfrac{k_q \Gamma^{(q)}}{b^q} \left(\dfrac{1 -r^q c_n}{1-c_n}\right) + o\left( \dfrac{1}{b^q}\right)\,,
\end{equation*}
where, setting $r > 1/c_n$ can again induce a positive first-order bias for positively correlated chains.

Choosing the truncation point $b$ crucially impacts the finite sample behaviour of the SV estimator.  The literature on optimal truncation points, which is also referred to as automatic bandwidth procedures, is far richer for SV estimators compared to optimal batch sizes for BM estimators.  An interested reader is directed to \cite{andr:1991}, \cite{chang2018understanding}, \cite{lazarus2021size}, \cite{lazarus2018har}, \cite{sun2013heteroskedasticity,sun2014let}, and \cite{sun2008optimal}. 

SV estimators offer greater flexibility compared to batching methods, as they provide a range of lag windows to choose from \citep[see e.g.][]{ande:1994}. However, SV estimators are less commonly utilized in practical MCMC simulations due to their slower calculation speeds. A potential remedy involves leveraging the efficiency of a fast Fourier transform to level the computation times. This strategy has been implemented in the \texttt{mcmcse} R package \citep{mcmcse2021} and will be employed for comparing computational efficiency in Section~\ref{sec:examples}.  

\section{Initial sequence estimators} \label{sec:initial}

\cite{geye:1992} introduced a conservative Monte Carlo error estimation method for a univariate mean, which was extended to multivariate settings by \cite{koso:2000} and \cite{dai:jone:2017}.  The fundamental concept involves leveraging the properties of the sequence of lag-$k$ covariance matrices to establish a truncation point for the truncated periodogram lag-window.   A more intricate shape-constrained estimator for univariate scenarios is considered by \cite{berg2022efficient}.  This class of estimators is appropriate for reversible Markov chains, i.e.\ those that satisfy detailed balance with respect to $F$.  Consequently, in this section we assume $R(k) = R(-k)$ for all integers $k$.

We confine our focus to multivariate initial sequence estimates of $\Sigma$ from \cite{dai:jone:2017}.  To this end, define the sum of an adjacent pair of covariances by $A_i = R(2i) + R(2i+1)$ for $i = 0, 1, 2, \dots$.  
Then $\Sigma$ can be rewritten as
\[
\Sigma = -R(0) + 2 \sum_{i=0}^{\infty} A_i
\]
and the $m$th partial sum can be defined as
\[
\Sigma_m = -R(0) + 2 \sum_{i=0}^{m} A_i .
\]
\cite{dai:jone:2017} show there exists a non-negative integer $m_0$ such that $\Sigma_m$ is positive definite for $m \ge m_0$ and not positive definite for $m < m_0$.  They go on to prove the sequence $\left\{ | \Sigma_m | : m_0, m_0 + 1, \dots \right\}$ is positive, increasing, and converges to $| \Sigma |$, which can be used to establish a truncation point in practice.

Recall the sample lag covariances at \eqref{eq:sample_lag_covariance} and define $\hat{A}_i =  \hat{R}(2i) + \hat{R}(2i+1)$ for $i = 0, 1, 2, \dots$.  Then the empirical estimator of $\Sigma_m$ for $0 \le m \le \lfloor n/2 - 1 \rfloor$ is
\[
\Sigma_{n,m} = -\hat{R}(0) + 2 \sum_{i=0}^{m} \hat{A}_i .
\]
Let $s_n$ be the smallest integer such that $\Sigma_{n,s_n}$ is positive definite, which we know exists when $n$ is sufficiently large.  Further, let $t_n$ be the largest integer $m \in \left\{ s_n , \dots, \lfloor n/2 - 1 \rfloor \right\} $ where $| \Sigma_{n,i} | > | \Sigma_{n,i-1} |$ for all $i \in \left\{ s_n +1, \dots, m \right\}$.  Then \cite{dai:jone:2017} define the multivariate initial sequence estimate as $\Sigma_{\text{seq},n} = \Sigma_{n, t_n}$.  The multivariate initial sequence estimate is conservative since $\liminf_{n\rightarrow\infty} |\Sigma_{\text{seq},n} | \ge | \Sigma |$ with probability 1. 

The estimator $\Sigma_{\text{seq},n}$ is constructed sequentially using the update
\[
\Sigma_{n, i+1} = \Sigma_{n, i} + 2 \hat{A}_{i+1}.
\]
Since $\hat{A}_{i+1}$ could potentially have negative eigenvalues, \cite{dai:jone:2017} suggest an adjusted update where negative eigenvalues are replaced by 0 at each step.  This process yields the adjusted multivariate initial sequence estimator, say $\Sigma_{\text{adj},n}$, which is also conservative in the sense that $\liminf_{n\rightarrow\infty} |\Sigma_{\text{adj},n} | \ge | \Sigma |$ with probability 1. 

Initial sequence estimators find limited use among MCMC practitioners due to their applicability only to reversible Markov chains.  Moreover, bias and variance properties of initial sequence estimators remain unknown.  Most critical to practitioners, initial sequence estimators also require greater computational effort.  We elaborate on these restrictions in the following section where all the calculations are readily available through the \texttt{mcse.initseq()} function within the \texttt{mcmcse} \texttt{R} package. 

\section{Example} \label{sec:examples}

We highlight differences of the BM, SV, and initial sequence estimators of $\Sigma$ using Bayesian logistic regression for modeling credit risk.  The dataset utilized is sourced from UC Irvine’s machine learning repository \citep{misc_statlog_(german_credit_data)_144}. The model we pursue aims to classify loan applicants into low and high credit risk groups using seven explanatory variables; status of existing checking account, credit history, duration, savings, other debts, housing status, and amount. 

Given that some explanatory variables are categorical, the proposed model contains 19 regression coefficients, including the intercept term. The Bayesian logistic regression model is represented as
\begin{equation*}
    P(Y_i = 1) = \frac{\exp \left( x_i^T \beta\right) }{1 + \exp \left( x_i^T \beta\right) }. 
\end{equation*}
We consider a multivariate normal prior distribution $\beta \sim N( 0 , \frac{1}{100}I_{19})$ and are interested in estimating posterior means for the 19 regression coefficients.  We sample from the resulting posterior using a Metropolis-Hastings sampler via the \texttt{MCMCpack} \texttt{R} package.  

Variability in the posterior mean estimates can be ascertained via estimation of an appropriate covariance matrix $\Sigma$, which can then be used to calculate MCSEs.  Instead, we consider ESS as a proxy for the MCSE because ESS is univariate and the quality of estimation of both MCSE and ESS are largely dependent on the estimation of $\Sigma$.  Specifically, we consider $\widehat{\text{ESS}}/n$ calculated using BM, SV, and initial sequence estimators of $\Sigma$ with their corresponding the original, zero, and over lugsail settings (when appropriate).  ESS was estimated using the \texttt{R} package \texttt{mcmcse} with BM, SV (with the Bartlett lag window), and initial sequence estimators.  Chain lengths considered are $n$ = 30k, 40k, \dots, 200k.  The simulation process was replicated 500 times for each estimator and chain size combination.

\begin{figure}
\begin{center}
 {\includegraphics[height=1.8in]{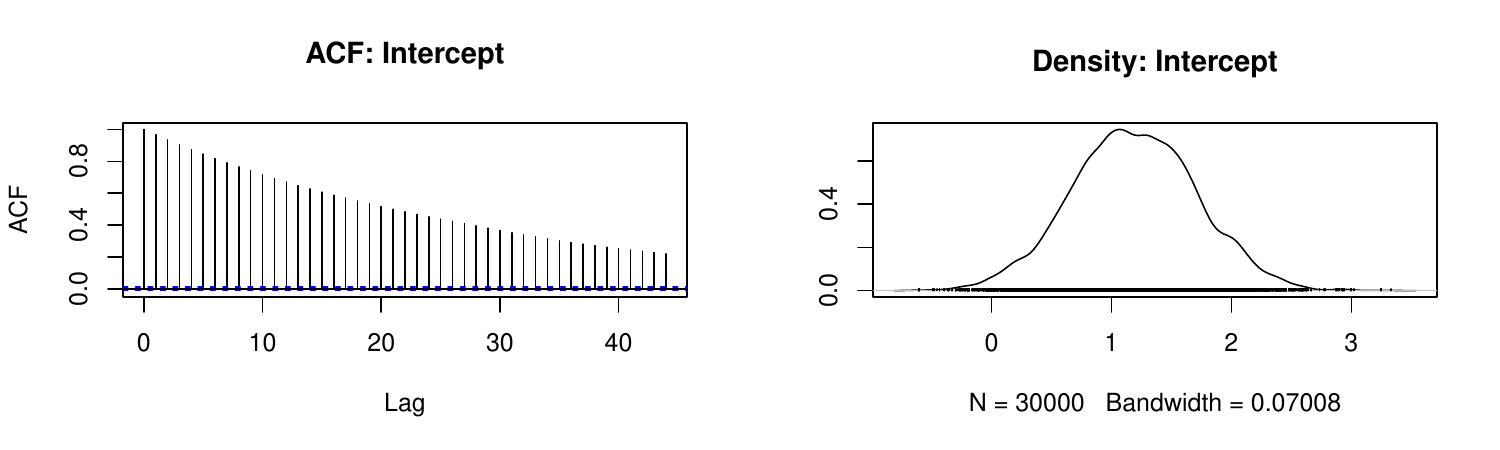}} 
 {\includegraphics[height=1.8in]{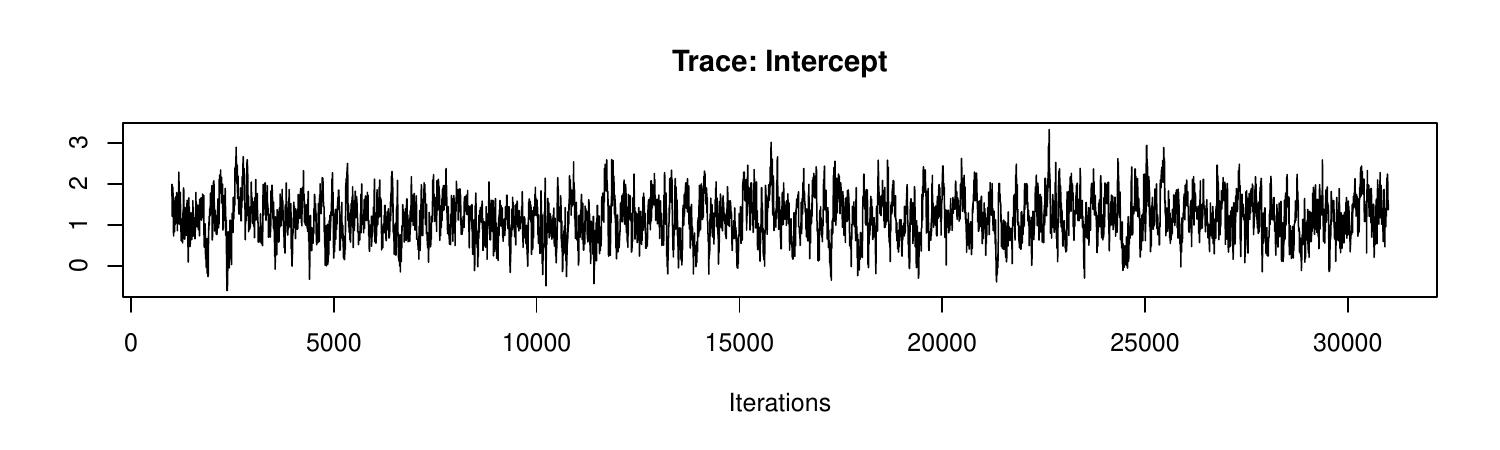}} 
 \caption{\label{fig:chain_mixing_intercept}Mixing properties of the the intercept coefficient.} 
\end{center}
\end{figure}

\begin{figure}
\begin{center}
 {\includegraphics[height=1.8in]{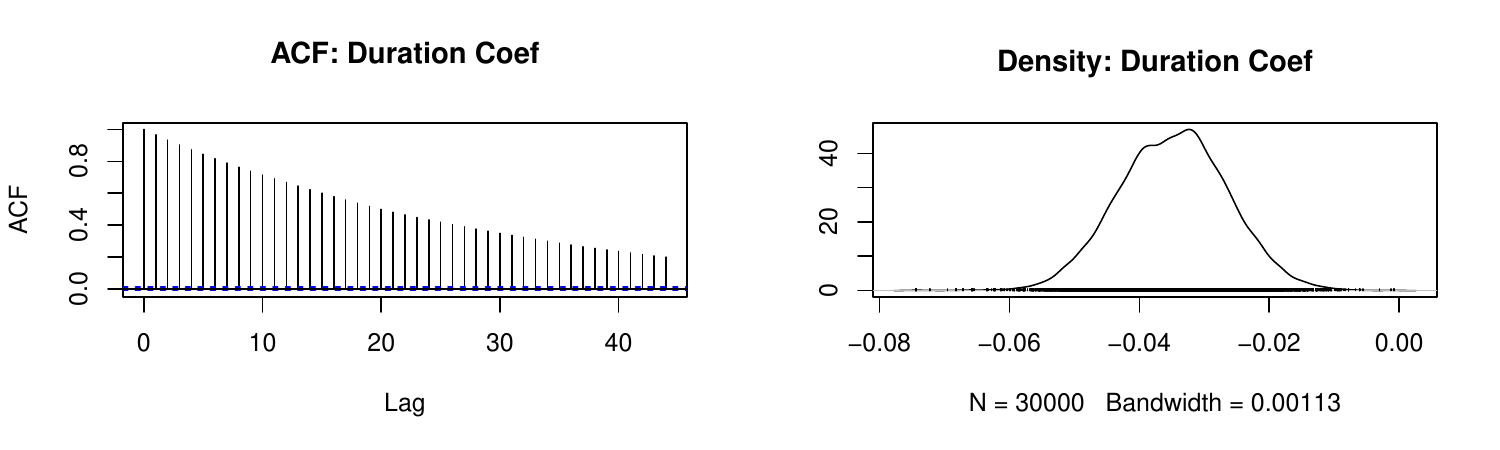}} 
 {\includegraphics[height=1.8in]{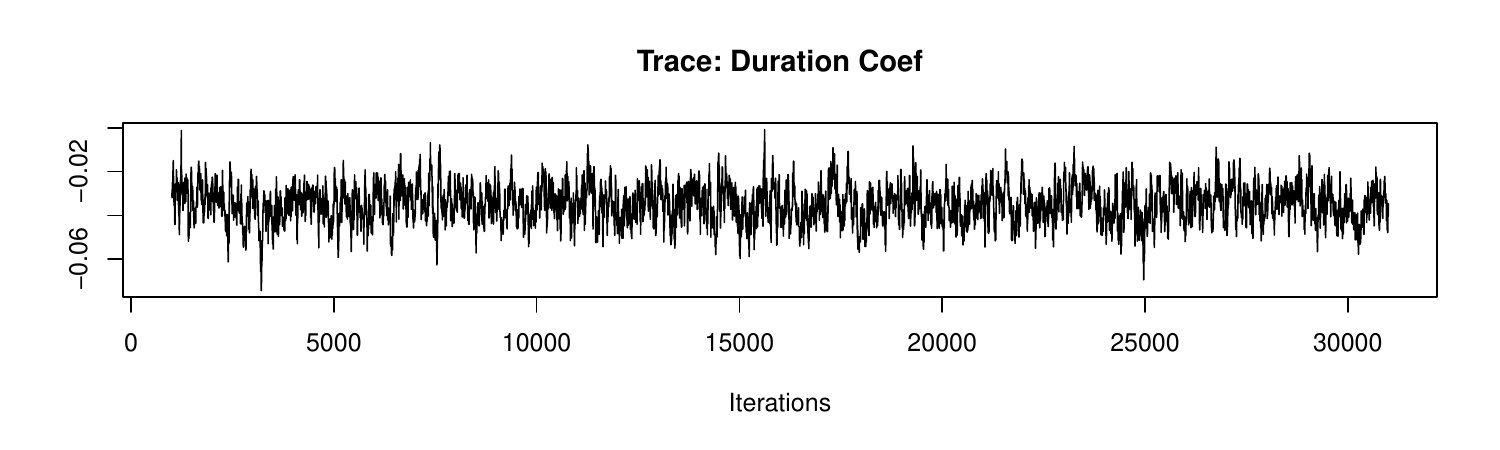}} 
 \caption{\label{fig:chain_mixing_coef}Mixing properties of the loan duration coefficient.}
\end{center}
\end{figure} 

It's impractical to display the mixing properties for all 19 coefficients in brevity.  Therefore, we concentrate on two coefficients for illustrative purposes, noting that the remaining coefficients demonstrate similar properties.  Figures~\ref{fig:chain_mixing_intercept} and~\ref{fig:chain_mixing_coef} illustrate strong correlation in the mixing properties for the intercept and loan duration coefficients.  Moreover, density plots exhibit reasonable patterns, suggesting appropriate mixing within the chain.

\begin{figure}
\begin{center}
 {\includegraphics[height=3in]{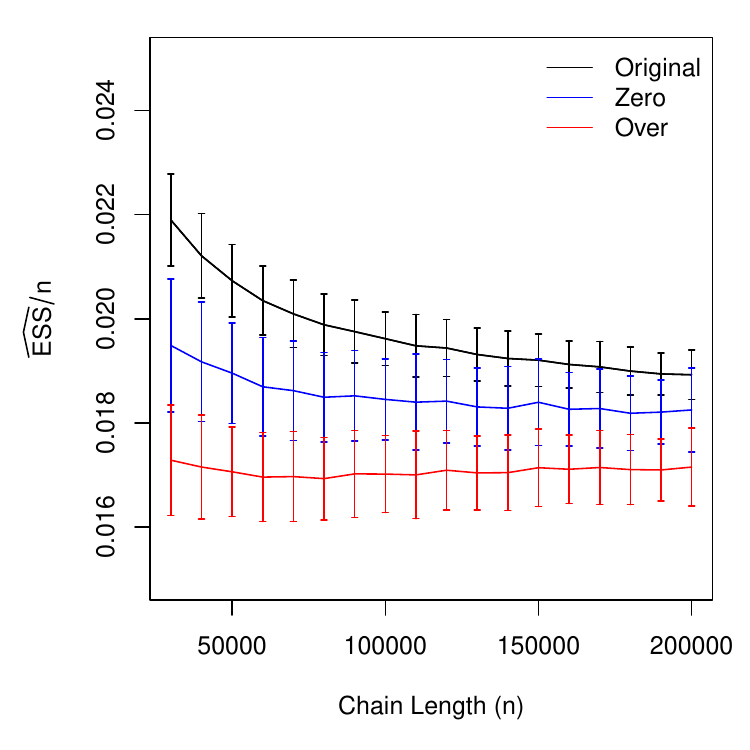}} 
 {\includegraphics[height=3in]{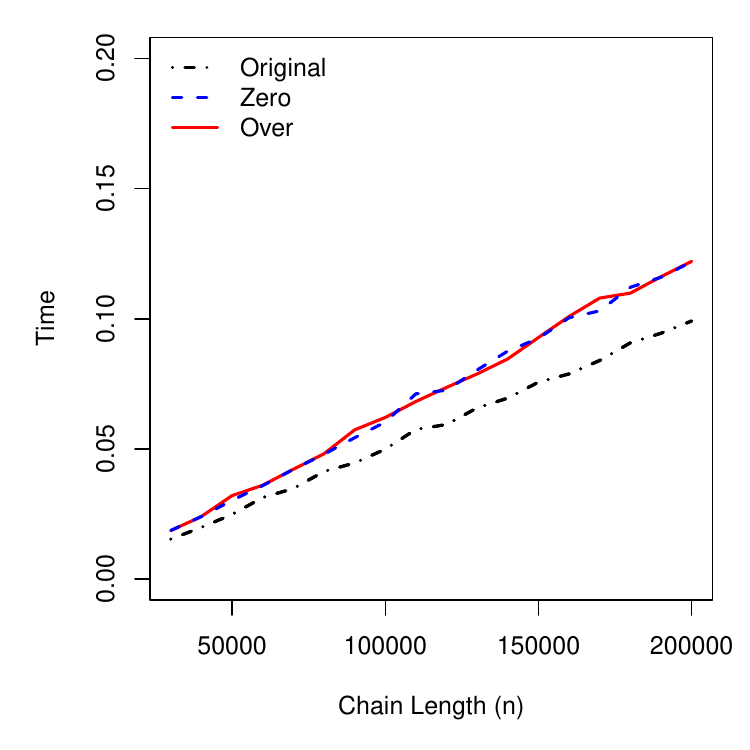}} 
 \caption{\label{fig:bm_example}Plots of $\widehat{\text{ESS}}/n$ and median computation time for the BM estimator.}
\end{center}
\end{figure} 

\begin{figure}
\begin{center}
 {\includegraphics[height=3in]{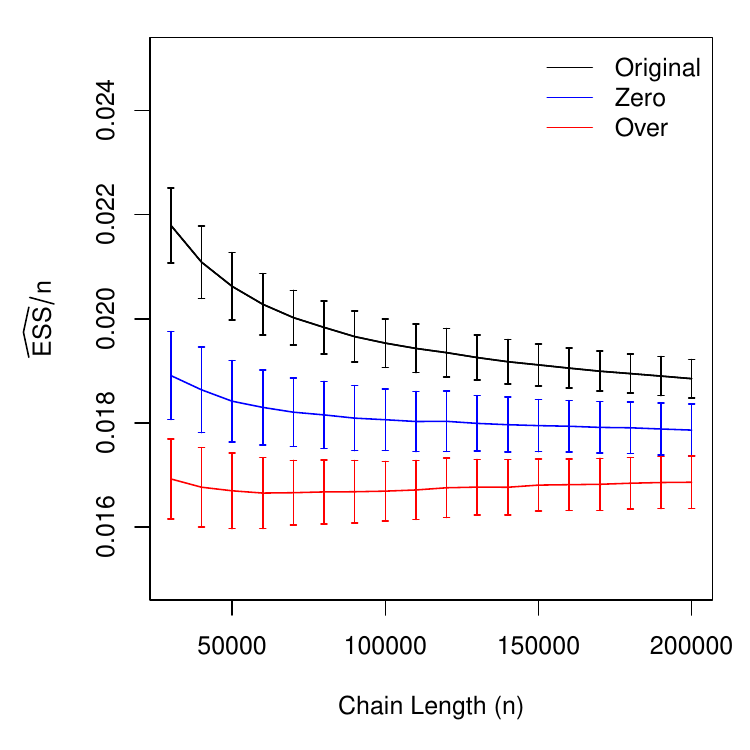}} 
 {\includegraphics[height=3in]{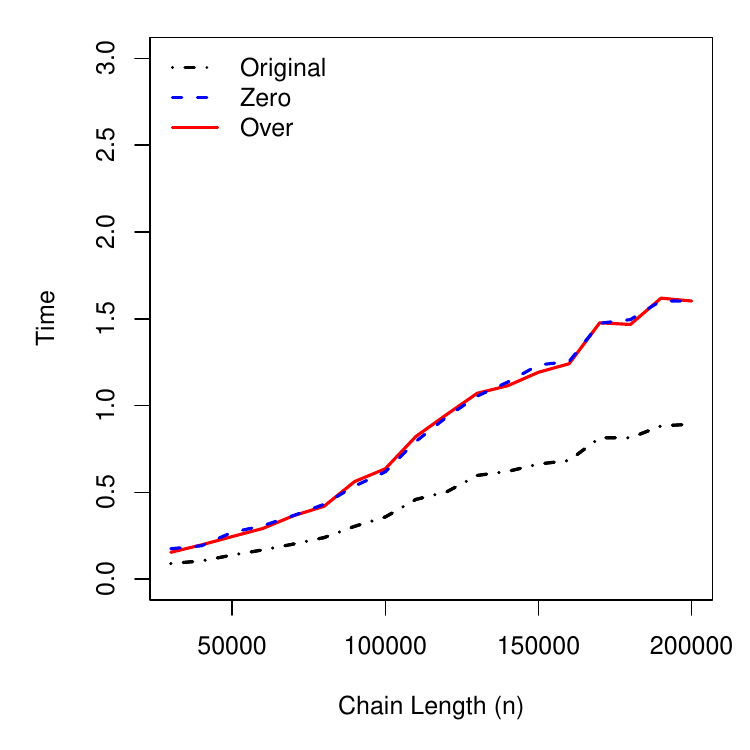}} 
 \caption{\label{fig:sv_example}Plots of $\widehat{\text{ESS}}/n$ and median computation time for the SV estimator.}
\end{center}
\end{figure}

\begin{figure}
\begin{center}
 {\includegraphics[height=3in]{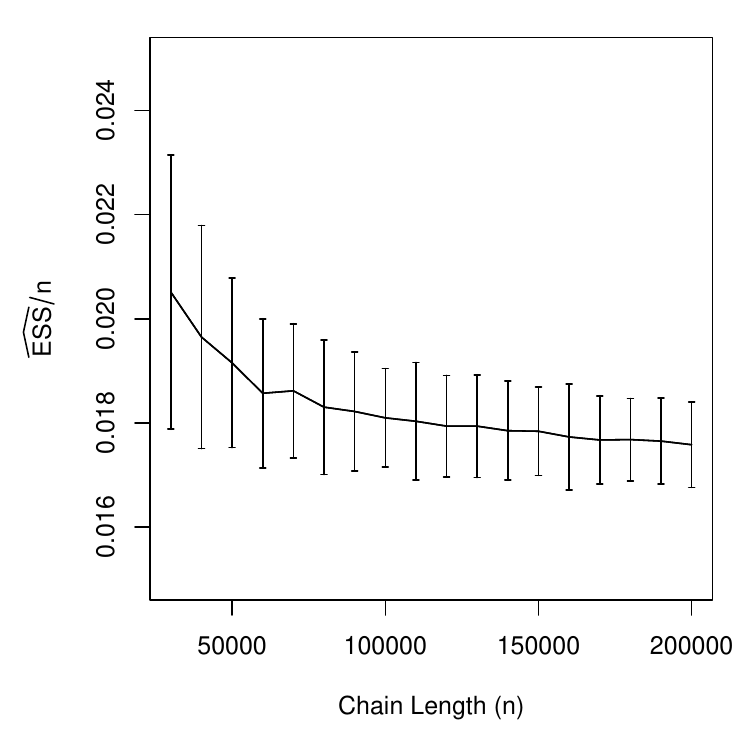}} 
 {\includegraphics[height=3in]{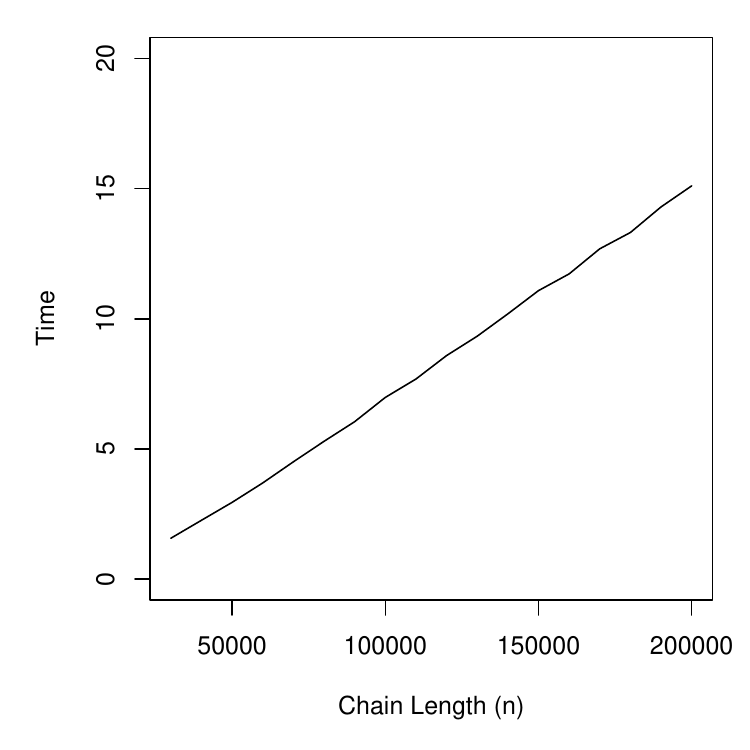}} 
 \caption{\label{fig:is_example}Plots of $\widehat{\text{ESS}}/n$ and median computation time for the initial sequence estimator.}
\end{center}
\end{figure}

The relationship between simulation size and and $\widehat{\text{ESS}}/n$ is illustrated on the left side of Figures~\ref{fig:bm_example}-\ref{fig:is_example} across the different estimation techniques.  The median computation time are also presented on the right side of Figures~\ref{fig:bm_example}-\ref{fig:is_example}.  To ease comparison, Tables~\ref{tab:avg_ess_estimate} and~\ref{tab:time_to_estimate} provide numerical estimates and compute times for the terminal chain length of 200k.  

Observe that $\Sigma_n$ is in the denominator for the estimator $\widehat{\text{ESS}}$ in \ref{eqn:ESS}, thus if $\Sigma_n$ is negatively biased $\widehat{\text{ESS}}$ can become inflated.  We observe this behavior in the left hand side of Figures~\ref{fig:bm_example}-\ref{fig:is_example} for the original SV and BM estimators, where $\widehat{\text{ESS}}$ converges from above and eventually stabilizes.  This convergence from above can result in premature chain termination when using the guidelines described in Section \ref{sec:output}. 

In contrast to the original estimators, the zero and over lugsail adjustments result in a less biased $\Sigma_n$. The consequences of this reduced bias is immediately observed via the faster rate of convergence for $\widehat{\text{ESS}}$. The over lugsail setting has the added feature that it typically converges from below, a characteristic procured due to a positively biased $\Sigma_n$. Thus, the over lugsail may result in running a chain slightly longer than necessary yielding a more favorable conservative approach to termination.  The initial sequence estimator also has an improved convergence rate in comparison to the original SV and BM estimators, but it does not converge as fast as the over lugsail and it does not seem to converge from above in this example. 

We further observe the error bars which indirectly indicate estimation variability of $\Sigma_n$.  As expected, the error bars are larger for the zero and over lugsail settings compared to the original estimator for both the SV and BM methods.  
In addition, the simulation illustrated a decrease in the variability of $\widehat{\text{ESS}}$ as the chain increases for all settings.  The difference of variability between the estimators is marginal, with the the initial sequence estimator is generally higher than the SV and BM methods. 

{\renewcommand{\arraystretch}{1.3} 
\begin{table}[ht]
\centering
\begin{tabular}{l|ccc}
  \hline
 & Original & Zero & Over \\ 
  \hline
Batch Means & 0.01892 (0.00029)& 0.01825 (0.00049) & 0.01715 (0.00045)  \\ 
Spectral Variance   &  0.01885 (0.00022)  & 0.01786 (0.00030)  & 0.01686 (0.00030)  \\ 
 Initial Sequence  &  0.01758 (0.00050)  &  & \\ 
   \hline
\end{tabular}
\caption{Average $\widehat{\text{ESS}}/n$ with a chain length of $n = 200k$.}
\label{tab:avg_ess_estimate}
\end{table}
}

{\renewcommand{\arraystretch}{1.3} 
\begin{table}[ht]
\centering
\begin{tabular}{l|ccc}
  \hline
 & Original & Zero & Over \\ 
  \hline
Batch Means & 0.09915  (0.00317)& 0.12180 (0.00474) & 0.12210 (0.00452) \\ 
Spectral Variance   &  0.8926 (0.09265) & 1.60100 (0.14400) & 1.60200 (0.13480) \\ 
 Initial Sequence  &  16.40000 (3.617) &  & \\ 
   \hline
\end{tabular}
\caption{Average time in seconds to calculate $\Sigma_n$ with a chain length of $n = 200k$}
\label{tab:time_to_estimate}
\end{table}
}

Finally, Table~\ref{tab:time_to_estimate} provides computation time across the estimation procedures for the terminal chain length of 200k.  BM methods clearly have a more favorable computation time and the initial sequence method had the slowest. For example, we notice in Table \ref{tab:time_to_estimate} the computation time for original BM and SV estimators roughly differed by a factor of about 10.  Moreover, the computation time between the SV and initial sequence estimators differed by over 15.  Implementing the lugsail adjustments increased computation time for a given estimator, with relatively minimal discrepancy between the zero and over lugsail settings. For this setting the computation times across all settings is in terms of seconds, and the difference in computation times is largely inconsequential.  However, with longer chain lengths and higher dimensions this difference will eventually become substantial. 

\section{Discussion} \label{sec:discussion}

We explored various techniques for estimating $\Sigma$, the asymptotic covariance matrix from a Markov chain CLT. Our focus lies on scenarios commonly encountered in contemporary high-dimensional MCMC simulations. These scenarios involve concurrent estimation of multiple quantities, which are likely correlated, and where the Markov chain correlation tends to be positive and strong. Moreover, the simulations involve an extensive number of iterations. Within these constraints, we identify diverse methods for estimating $\Sigma$ beyond the prevalent BM estimator. 

For other scenarios with correlated data, variance estimators often follow a structure akin to $\Sigma$. For instance, in time series analysis $\Sigma$ emerges in the context of spectral estimation and the determination of long-run variance \citep{hannan:1970,priest:1981}.  In econometrics, it is referred to as heteroskedastic and autocorrelation consistent covariance matrix estimation \citep{andr:1991, newey:west:1987}.  While the literature concerning the estimation of $\Sigma$ in time-series and econometric applications is still expanding, these contexts typically revolve around moderate levels of correlation, a relatively limited number of available data points for estimation purposes, and smaller number of dimensions, see e.g.\ \cite{lazarus2018har}.

It is essential to underscore that the estimation of the LRV in correlated data scenarios continues to be an active research area.  \cite{chan:yau:2017} propose a recursive estimator using a sequence of batch sizes that increase with $n$, while \cite{alexopoulos2007overlapping} considers overlapping batching methods for steady-state simulation output.  Other general techniques include standardized time series weighted area estimators \citep{goldsman1990properties} and subsampling bootstrap variance estimators \citep{poli:roma:wolf:1999}.  

Specialized approaches include estimates of $\Sigma$ based on regenerative simulations \citep{hobe:jone:pres:rose:2002}, solutions of the Poisson equation \citep{douc2022solving}, and shape-constrained estimators based on underlying restrictions similar to initial sequence estimation \citep{berg2022efficient}.  Additionally, \cite{mcelroy2024estimating} address LRV estimation through local polynomial regression on the periodogram of spectral density function at zero. Estimation of the asymptotic covariance matrix in nonstationary time series has also been addressed in the presence of change points \citep{chan2022optimal, chan2022mean}.  Nonetheless, estimating $\Sigma$ in truly high-dimensional problems remains challenging.

A practitioner may be tempted to select the widely used BM method without much consideration of the setting. However, all the discussed estimation methods are valid and should be considered. The choice of estimator revolves around underlying correlation, computational costs, and downstream analysis, particularly estimating the ESS and evaluating parameter MCSEs to create confidence intervals.

Lugsail transformations for both BM and SV estimators are more effective in the presence of high correlation, preventing premature termination with only a marginal increase in computational burden. However, over lugsails may over inflate the MCSE for the parameters of interest. Compared to the BM method, SV estimation increases computational burden but offers only a marginal improvement in MCSE and ESS estimation. The initial sequence estimator serves a similar purpose but involves a substantial increase in computational complexity and minimal tuning parameter adjustment.

Each method offers unique advantages, and a definitive choice is not always apparent.  Therefore, the objective of $\Sigma$ estimation is less about achieving optimality and more about making well-informed decisions. 

\section*{Acknowledgments}
The authors would like to thank the editorial team and referee for their feedback that led to significant improvements in our manuscript.

\bibliographystyle{apalike} 
\bibliography{ref.bib}

\end{document}